\documentclass[aps,prl,twocolumn,superscriptaddress,groupedaddress,preprintnumbers]{revtex4-1} 
\usepackage{graphicx}  
\usepackage{dcolumn}   
\usepackage{bm}        
\usepackage{amssymb}   
\usepackage{amsmath}
\usepackage{mathtools}
\usepackage{cases}
\usepackage{mathrsfs}
\usepackage{color}
\usepackage{CJK}
\usepackage{ulem}

\newcommand{\be}{\begin{equation}}
\newcommand{\ee}{\end{equation}}
\newcommand{\bea}{\begin{eqnarray}}
\newcommand{\eea}{\end{eqnarray}}

\begin{document}
\begin{CJK*}{GB}{}
	\title{Reconstruction of Power Spectrum of Primordial Curvature Perturbations on small scales from Primordial Black Hole Binaries scenario of LIGO/VIRGO detection}
	\author{Xinpeng Wang${}^{a}$,~Ying-li Zhang${}^{a,b,c,d}$,~Rampei Kimura{}$^{e}$~and Masahide Yamaguchi${}^{f}$\\
		\it
		$^{a}$School of Physics Science and Engineering, Tongji University, Shanghai 200092, China\\
		$^{b}$Institute for Advanced Study of Tongji University, Shanghai 200092, China\\
		$^{c}$Kavli Institute for the Physics and Mathematics of the Universe (WPI), Chiba 277-8583, Japan\\
		$^{d}$Center for Gravitation and Cosmology, Yangzhou University, Yangzhou 225009, China\\
		$^{e}$Waseda Institute for Advanced Study, Waseda University, 1-6-1 Nishi-Waseda, Shinjuku, Tokyo 169-8050, Japan\\
		$^{f}$Department of Physics, Tokyo Institute of Technology, 2-12-1 Ookayama, Meguro-ku, Tokyo 152-8551, Japan\\}
	    
	\date{\today}
	\begin{abstract}
As a candidate bound for the Binary Black Hole (BBH) merger events detected by LIGO/Virgo, Primordial Black Holes (PBHs) provide a useful tool to investigate the primordial curvature perturbations on small scales. Using the GWTC-1 to GWTC-3 catalogs, under the scenario that PBHs originate from large primordial curvature perturbations on small scales during the inflationary epoch, we for the first time reconstruct the power spectrum of primordial curvature perturbations on small scales. It is found that the value of the amplitude of the primordial power spectrum is enhanced to $\mathcal{O}\left(10^{-2}\right)$ on scales $\mathcal{O}\left(1\right){\rm pc}$. This may imply the validity of PBH as a possible BBH merger candidate.
\end{abstract}
	\maketitle
	\end{CJK*}

\textit{Introduction.}~
The O1 to O3 runs of the Laser Interferometer Gravitational-wave Observatory (LIGO) Collaboration and Virgo Collaboration~\cite{Abbott:2016blz,Abbott:2016nmj,TheLIGOScientific:2016pea,Abbott:2017vtc,Abbott:2017oio,Abbott:2017gyy,LIGOScientific:2018mvr,Venumadhav:2019lyq,Abbott:2020niy,LIGOScientific:2021usb,LIGOScientific:2021djp} have revealed two unexpected characteristics of most of the detected binary black hole (BBH) mergers: the high masses (larger than 20 $M_\odot$) and low effective spin. The origin of these heavy BHs and the formation of such BBHs which merge within the age of the universe is still in debate~\cite{Belczynski:2010tb,Dominik:2012kk,Dominik:2013tma,Dominik:2014yma,Belczynski:2016obo,TheLIGOScientific:2016htt,Miller:2016krr}. One of the fascinating possible explanations is presented by the Primordial Black Holes (PBHs) which could form from the collapse of large density fluctuations in the early stages of the universe~\cite{Bird:2016dcv,Clesse:2016vqa,Sasaki:2016jop,Sasaki:2018dmp, Garcia-Bellido:2020pwq}. Typically, if the power spectrum of primordial curvature perturbation has a peak on some small scale, at the horizon reentry, when the density perturbation of the corresponding regions
exceeds the threshold value $\Delta_\text{th}$, the matter inside the Hubble horizon will collapse and form a PBH~\cite{Zeldovich:1963,Hawking:1971ei,Carr:1974nx,Meszaros:1974tb,Carr:1975qj}. The PBH mass is of the same order of the total energy contained inside the Hubble radius at horizon reentry, whose length scale corresponds to the inverse of the wavenumber of the peak.

The PBH formation scenario does not violate the current observational constraint because the amplitude of the primordial curvature perturbation is tightly constrained only on the cosmic microwave background (CMB) scales~\cite{Bringmann:2011ut,Green:2018akb,Planck:2018jri}. On small scales (smaller than Mpc scale), although some upper bounds have been obtained~\cite{Chluba:2019kpb,Jeong:2014gna,Nakama:2014vla,Inomata:2016uip,Allahverdi:2020bys,Gow:2020bzo,Franciolini:2022pav,Franciolini:2022tfm}, we still lack the information of primordial curvature perturbations. This situation may be significantly improved if (some of) the detected BBH mergers could be attributed to PBHs. Under this expectation, a method to systematically reconstruct the power spectrum on small scales has been proposed in~\cite{Kimura:2021sqz}. The main idea is to formulate the relation among the merger rate $\mathcal{R}(m_1, m_2, t)$, the PBH mass function $f(m)$ and the power spectrum $\mathcal{P}_{\mathcal{R}}(k)$, then link them together to find a one-to-one correspondence between the merger rate and the power spectrum. With the increasing number of the detected BBH merger events, especially the data from O3 run of LIGO/Virgo, in this letter, we furthermore generalize this method and apply it to reconstruct the power spectrum of primordial curvature perturbation on scale $\mathcal{O}$(1) pc for the first time. \\

\textit{The method of reconstruction.}~
Here we briefly describe the method of reconstruction. When we have known the PBH binary merger rate, the power spectrum of primordial curvature perturbations can be reconstructed by the following steps:
\begin{align}\label{chain}
\mathcal{R}(m_1, m_2, t)\rightarrow f(m)\rightarrow\sigma^2(R)\rightarrow\mathcal{P}_{\mathcal{R}}(k)\,,
\end{align}
where $\mathcal{R}(m_1, m_2, t)$ is the merger rate density of the PBH binaries with individual masses $m_1$ and $m_2$ at cosmic time $t$, $f(m)$ is the PBH mass function, $\sigma^2(R)$ is the variance of the density perturbation smoothed over a comoving length scale $R$, and $\mathcal{P}_{\mathcal{R}}(k)$ is the power spectrum of the primordial curvature perturbations. Since the difference in the coalescence time $t$ for individual BBH merger events is small compared to the age of the universe, we neglect the time dependence of merger rate and express it as $\mathcal{R}(m_1, m_2)$. The one-to-one relations of these key qualities can be obtained analytically in each step under the assumptions:
(1). PBHs form out of rare high-$\sigma$ peaks of the primordial curvature perturbations in the radiation dominated era; (2). The window function takes the {\it top-hat} form in $k$-space; (3). The primordial curvature perturbation follows {\it Gaussian} probability distribution. The main steps of the method are as follows:

Firstly, from the expression of merger rate of PBHs, neglecting the suppression factor, its mass dependence can be expressed as~\cite{Raidal:2018bbj}
\begin{align}
\label{ff}
 f\left(m_{1}\right)f\left(m_{2}\right)\propto\left({m_{1}+m_{2}}\right)^{-\frac{36}{37}}(m_1m_2)^{-\frac{3}{37}}\mathcal{R}\left(m_{1}, m_{2}\right)\,.
\end{align}
A simple recurrence relation of mass function is deduced in~\cite{Kimura:2021sqz} where only the merger rates with equal individual mass are relevant. However, in practice, due to the limited number of observed PBH merger events, it is necessary to generalize the recurrence relation to include the merger rate of different individual masses. From \eqref{ff}, we obtain the following relation:
\begin{align}
    f\left(m_{2}\right)&=f\left(m_{1}\right)\frac{\mathcal{R}\left(m_{1}, m_{2}\right)}{\mathcal{R}\left(m_{1}, m_{1}\right)}\left(\frac{m_{1}}{m_{2}}\right)^{\frac{3}{37}}\left(\frac{2m_{1}}{m_{1}+m_{2}}\right)^{\frac{36}{37}}\,,\label{f12}\\
    f\left(m_{n}\right)&=f\left(m_{n-2}\right)\frac{\mathcal{R}\left(m_{n-1}, m_{n}\right)}{\mathcal{R}\left(m_{n-2}, m_{n-1}\right)}\left(\frac{m_{n-2}}{m_{n}}\right)^{\frac{3}{37}}\times\nonumber\\
    &\qquad\times\left(\frac{m_{n-2}+m_{n-1}}{m_{n}+m_{n-1}}\right)^{\frac{36}{37}}\,, \quad n\geqslant3\,, \label{fn} 
\end{align}
where $m_n=m_1+(n-1)\Delta m$.

Provided with the merge rate from LIGO/Virgo detection, setting an initial value of $f(m_1)$ fixed by the normalization condition $\sum_n f(m_n) \Delta m = 1$, the value of $f(m_2)$ can be obtained from \eqref{f12}, and so on. We note that there exists a consistency relation for the PBH scenario
\begin{align}\label{Rm1m2}
\mathcal{R}(m_1, m_2)&=\left(\frac{m_1+m_2}{2\sqrt{m_1m_2}}\right)^{\frac{36}{37}}\sqrt{\mathcal{R}(m_1, m_1)\mathcal{R}(m_2, m_2)}\,.
\end{align}
which will be used to justify the mass range of the reconstruction procedure in the next section.

Secondly, under the assumption (3), the distribution of density contrast $\Delta$ is expressed as $P(\Delta)=e^{-(\Delta^2/2\sigma^2)}/\sqrt{2\pi\sigma^2}$ with the variance $\sigma^2$. In the non-critical collapse case, $\sigma^2$ is expressed in terms of the mass function $f(m)$ by using the Press-Schechter approach \cite{Kimura:2021sqz}:
\begin{align}\label{sigmafun1}
\sigma(m(R))=\frac{\Delta_{\rm th}}{\sqrt{2}}\left[{\rm erfc}^{-1}\left(\frac{2f_{\rm PBH}\Omega_{\rm CDM}}{\left(M_{\rm eq}K^3\right)^{\frac12}}m^{\frac{3}{2}}f(m)\right)\right]^{-1}\,,
\end{align}
where $f_{\rm PBH}\equiv\Omega_{\rm PBH}/\Omega_{\rm DM}$ is the energy fraction of PBHs in dark matter, $M_{\rm eq}$ is the horizon mass at the matter-radiation equality epoch, $K$ is the ratio between the mass of PBH and horizon mass.

Thirdly, the variance $\sigma^2$ is related to the mean value of the square of the density contrast smoothed over the comoving scale $R$ 
\begin{align}\label{sig}
\sigma^2 (R)=\langle \Delta_R^2 (t_R,{\bf x}) \rangle =\int_0^\infty W^2(kR) \mathcal{P}_\Delta(t_R,k)~\mathrm{d}(\ln k).
\end{align}
Under the assumption (2) of a top-hat form window function  
\begin{align}\label{hat}
W(kR)=\left\{\begin{matrix}
\displaystyle \quad 1\quad\,; & 0<k<1/R\,,&~\\
\\
\displaystyle \quad 0\quad; & \mathrm{otherwise}\,,
\end{matrix}
\right.
\end{align}
since the power spectrum of density contrast $\mathcal{P}_\Delta$ is related to that of curvature perturbation during the radiation-dominated epoch such that $\mathcal{P}_\Delta(t, k)=16k^4\mathcal{P}_{\mathcal{R}}(k)/(81a^4H^4)$, inverting \eqref{sig}, we obtain
\begin{align}\label{powersim}
\mathcal{P}_{\cal R} (k)&=\frac{81}{16}\left(4\sigma^2+k\frac{{\rm d}\sigma^2}{{\rm d}k}\right)\bigg|_{R=1/k}\,.
\end{align}

Principally speaking, the power spectrum is determined from the merger rate of PBH binaries by linking these three steps together. We will explain how we apply this method in practice.\\

\begin{figure} 
	\includegraphics[width=0.5\textwidth]{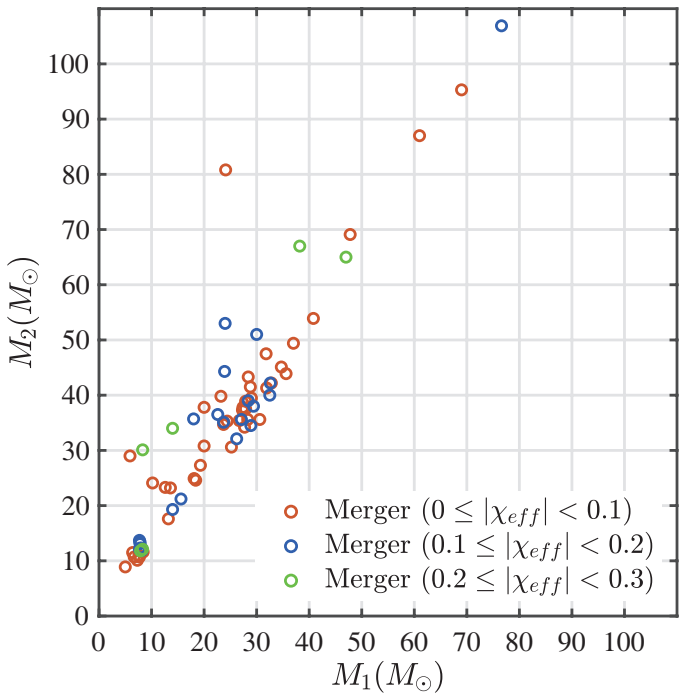}
	\caption{The distribution of 73 BBH merger events collected from the GWTC-1 to
 GWTC-3 catalogs under the requirement $m\gtrsim3M_{\odot}$ and $|\chi_{\rm eff}|\lesssim0.3$.}
\label{fig:allPBH} 
\end{figure}

\textit{Inclusion of LIGO/Virgo Data}~
We firstly pick up the observed events of BBH mergers relevant to the PBH scenario. In order to capture the idea that PBHs produced during the radiation-dominated era are expected to be one of the candidates for the detected BBH mergers with high mass and low spin~\cite{DeLuca:2020bjf}, we choose the BBH merger events from the GWTC-1 to GWTC-3 catalogs~\cite{LIGOScientific:2021usb,LIGOScientific:2021djp} with individual mass $m\gtrsim3M_{\odot}$ to exclude the possibility of neutron stars, and effective spin $|\chi_{\rm eff}|\lesssim0.3$ to collect BBHs with low effective spins. Totally, there are 73 events satisfying the conditions. We identify all of them as PBH mergers. They are plotted in the $M_1-M_2$ plane in Fig.~\ref{fig:allPBH} under the arrangement $M_2>M_1$ for definiteness.

We express the merger rate inferred from the LIGO/Virgo detection as
\begin{align}\label{Rdef}
\mathcal{R}\left(m_{i}, m_{j}\right)=\mathcal{R}_0N(m_{i}, m_{j})\,,\quad \mathcal{R}_0\equiv \frac{R_{\rm{total}}}{N_{\rm{total}}}\,,
\end{align}
 where $N(m_{i}, m_{j})$ is the number of PBH merger events included in the square grid which spans in the range $M_1\in(m_i-\frac{\Delta m}{2}, m_i+\frac{\Delta m}{2})$ and  $M_2\in(m_j-\frac{\Delta m}{2}, m_j+\frac{\Delta m}{2})$, respectively.  $\mathcal{R}_0$ is the average value of merger rate where $R_{\rm{total}}\equiv\sum_{i, j}\mathcal{R}\left(m_{i}, m_{j}\right)$ is the total merger rate implied by LIGO/Virgo, with dimension $\left({\rm Gpc}\right)^{-3}\cdot{\rm yr}^{-1}$, while $N_{\rm{total}}\equiv\sum_{i, j}N(m_{i}, m_{j})$ is the total number of individual 
PBH merger events. Hence, $\mathcal{R}\left(m_{i}, m_{j}\right)$ in \eqref{f12}--\eqref{Rm1m2} is replaced by the number of PBH events $N\left(m_{i}, m_{j}\right)$. 

\begin{figure}
	\includegraphics[width=0.5\textwidth]{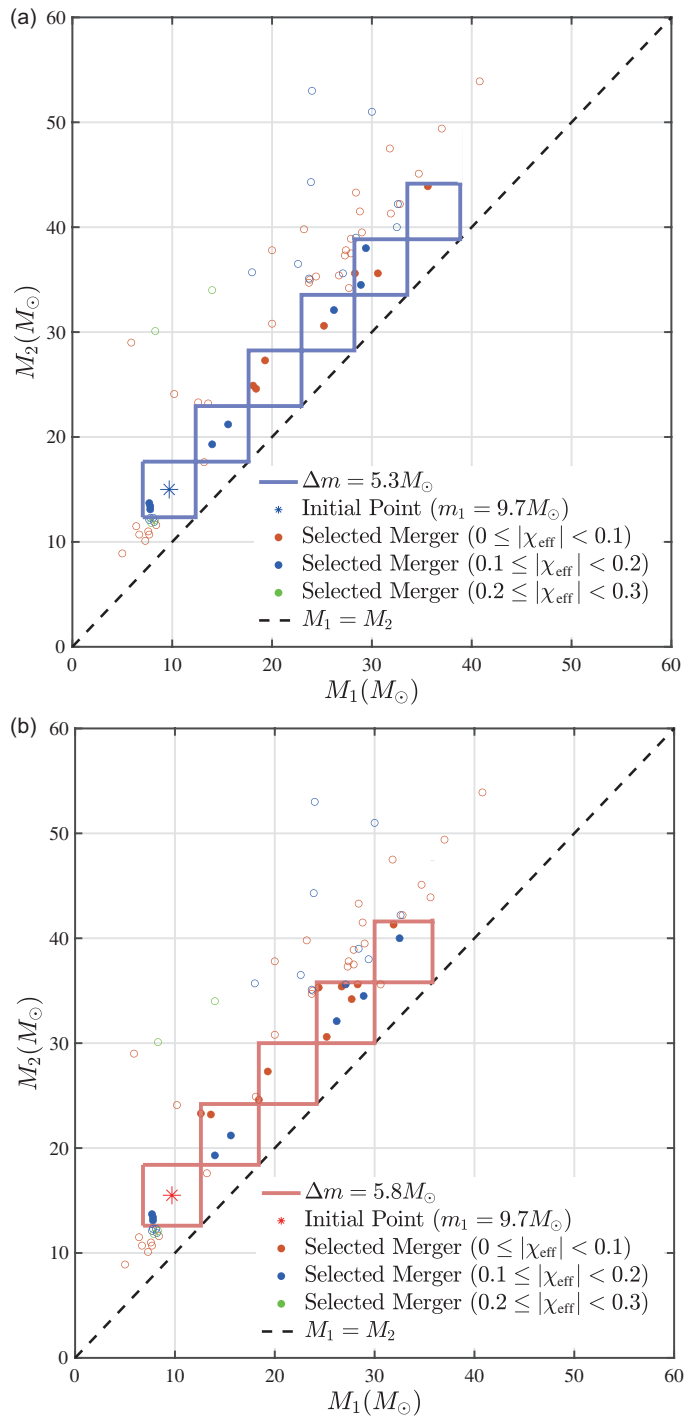}
	\caption{The PBH merger event number $N(m_{n-1}, m_n)$ is identified as the number of PBH merger events with individual mass in the range $(m_{n-1}-\frac{\Delta m}{2}, m_{n-1}+\frac{\Delta m}{2})$ and  $(m_n-\frac{\Delta m}{2}, m_n+\frac{\Delta m}{2})$, respectively. The mass grids on the $M_1-M_2$ plane are taken in the way $m_n=m_1+(n-1)\Delta m$ with mass gap $\Delta m$. We take the initial point $m_1=9.7M_{\odot}$, while the mass gap $\Delta m$ is taken as $5.3M_{\odot}$ (upper panel) and $5.8M_{\odot}$ (lower panel), respectively. Correspondingly, this gives 6 (upper panel) and 5 (lower panel) effective grids, respectively.}
	\label{fig:grids} 
\end{figure}

\begin{figure}
	\includegraphics[width=0.5\textwidth]{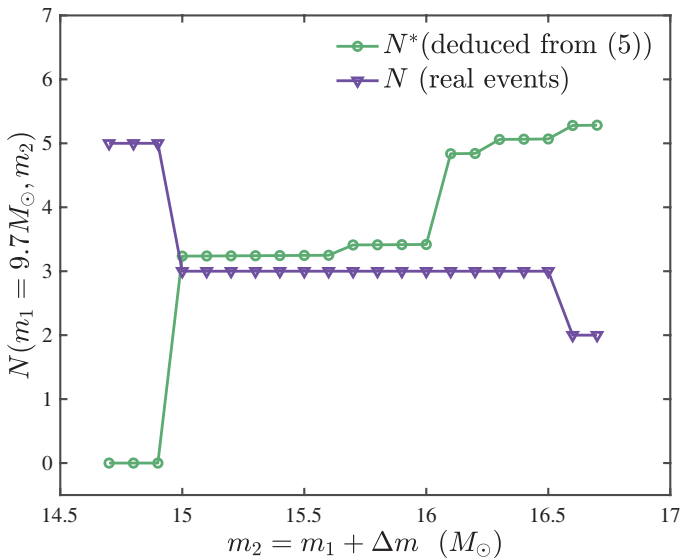}
	\caption{The comparison of the ``theoretical" number of PBH events $N^*$ deduced from the consistency relation \eqref{Rm1m2} to the detected event number $N$. We find that when $m_1=9.7M_{\odot}$, $\Delta m\in\left[5.3M_{\odot}, 5.8M_{\odot}\right]$, $N^*$ varies from 3.236 to 3.245. This best fits the detected event number $N(m_1, m_1+\Delta m )=3$. This infers the values of $m_1$ and $\Delta m$ in Fig.~\ref{fig:grids}}
\label{fig:consistency} 
\end{figure}

\begin{figure}
	\includegraphics[width=0.48\textwidth]{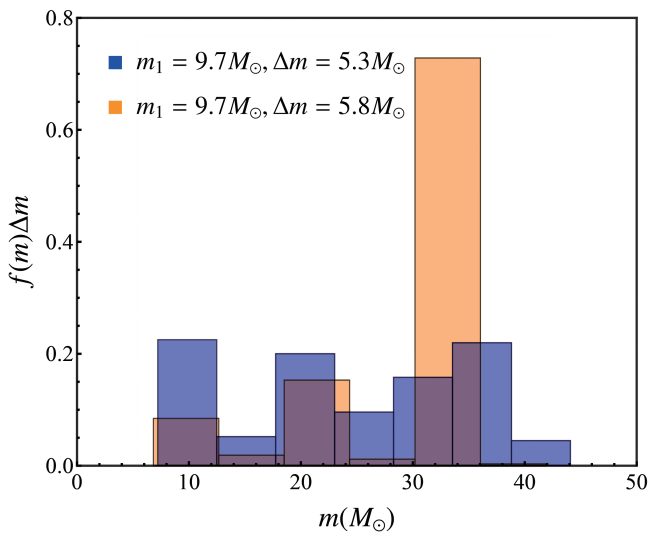}
	\caption{The values of dimensionless mass function $f(m_n) \Delta m$ obtained from \eqref{f12}--\eqref{fn} under the normalization condition $\sum_n f(m_n) \Delta m = 1$. In both cases we take the initial mass $m_1=9.7M_{\odot}$. The blue and orange columns correspond to $\Delta m=5.3M_{\odot}$ and $5.8M_{\odot}$, respectively.} 
	\label{fig:mfm} 
\end{figure}

Although currently there is no definite way to justify the range of the values of the initial mass $m_1$ and mass gap $\Delta m$, the consistency relation \eqref{Rm1m2} plays the role of a necessary condition to determine the range of values. Denoting $N^*(m_1, m_1+\Delta m)$ as the theoretically effective number of events deduced from \eqref{Rm1m2}, we found that when $m_1=9.7M_{\odot}$, $\Delta m\in\left[5.3M_{\odot}, 5.8M_{\odot}\right]$,  $N^*$ varies from 3.236 to 3.245, which gives the closest fit to the detected number 3 in the grid, as shown in Fig.~\ref{fig:consistency}. 
Hence, we present two typical cases where the initial point $m_1=9.7M_{\odot}$ while the mass gap $\Delta m$ is taken as $5.3M_{\odot}$ and $5.8M_{\odot}$, respectively. Since the merge rate involved in \eqref{f12} and \eqref{fn} has the form $\mathcal{R}\left(m_{n-1}, m_{n}\right)$ except for the initial point $\mathcal{R}\left(m_{1}, m_{1}\right)$, the grids are arranged in the way that the central point of each grid lies on the line $m_n=m_1+(n-1)\Delta m$. As shown in Fig~\ref{fig:grids}, this gives 6 (upper panel) and 5 (lower panel) effective grids, respectively, each grid contains nonzero number of PBH merger events. Counting from the left-hand side, the number of PBH merger events $N$ contained in the grids are 3, 2, 3, 2, 4, 1 for $\Delta m=5.3M_{\odot}$ case and 3, 4, 2, 8, 2 for $\Delta m=5.8M_{\odot}$ case, respectively. Both cases give $N\left(m_{1}, m_{1}\right)=10$ which is involved in \eqref{f12}.
Inserting the information of merger rate into \eqref{f12}--\eqref{fn}, in case of $\Delta m=5.3M_{\odot}$, we totally obtain 7 discrete values for the mass function: 6 of them are from $f(m_2)$ to $f(m_7)$, each of which is expressed in terms of $f(m_1)$, while $f(m_1)$ is obtained from the normalization condition $\sum_n f(m_n) \Delta m = 1$, as shown in Fig.~\ref{fig:mfm}. We notice that $f(m_n)$ obtained here should be understood as the average value of mass function which is averaged in the mass range $(m_n-\frac{\Delta m}{2}, m_n+\frac{\Delta m}{2})$. Similarly, in $\Delta m=5.8M_{\odot}$ case, 6 discrete values for the mass function are obtained. It is observed that in this case, there is a high peak at $m_5=32.9M_{\odot}$ (the fifth column in orange color counted from the left-hand side in Fig.~\ref{fig:mfm}). This is because (i). 22 BBH merger events concentrate in the mass range $M_1\in[20M_{\odot}, 30M_{\odot}]$ and $M_2\in[30M_{\odot}, 40M_{\odot}]$; (ii). the forth grid is mainly included in this region on the $M_1-M_2$ plane (see the lower panel of Fig.~\ref{fig:grids} ). These two facts lead to a large increase in $N\left(m_{4}, m_{5}\right)$ compared to $N\left(m_{3}, m_{4}\right)$, hence result in the high peak in $f(m_5)$ obtained from \eqref{fn}. 

Having obtained the mass function, although it is straightforward to obtain the variance $\sigma^2(m)$ from \eqref{sigmafun1}, uncertainties come from the values for $\Delta_{\rm th}$ and $f_{\rm PBH}$. Depending on the profile of perturbations, the threshold value of the comoving density contrast could vary from 0.2 to 0.6~\cite{Musco:2004ak,Musco:2008hv,Musco:2012au,Harada:2013epa,Yoo:2018kvb,Musco:2018rwt}. In the optimistic case, the critical peak value of curvature perturbation $\zeta_c\approx0.52$ is deduced from the peak theory for the monochromatic power spectrum~\cite{Yoo:2018kvb}. In the Press-Schechter formalism, this corresponds to $\Delta_{\rm th}\approx0.23$ for PBH formation during the radiation-dominance epoch. Meanwhile, the effect of PBHs accretion on CMB frequency spectrum and angular temperature/polarization power spectra constrains $f_{\rm PBH}\lesssim10^{-3}$ for PBH mass $\gtrsim10M_\odot$~\cite{Ali-Haimoud:2016mbv, Poulin:2017bwe, Serpico:2020ehh}. Hence, we take $\Delta_{\rm th}=0.23$ and $f_{\rm PBH}=10^{-3}$ for definiteness. Inserting these values, together with $M_{\rm eq}=3.52 \times 10^{17} M_{\odot}$ into \eqref{sigmafun1}, the variance $\sigma^2(m)$ is determined uniquely.
In the non-critical collapse case, the mass of PBH is approximated as the fraction of horizon mass $M_R$ such that $m=KM_R$ with $K\approx0.2$~\cite{Carr:1975qj}, so the mass of PBH is related to the horizon scale $k=1/R$ as~\cite{Green:2004wb}  
\begin{align}
\label{massk}
    m(k)=\left(\frac{k_{\mathrm{eq}}}{k}\right)^{2} M_{\mathrm{eq}} K\left(\frac{g_{*, \mathrm{eq}}}{g_{*}}\right)^{\frac{1}{3}},
\end{align}
where $g_*$ is the number of relativistic degrees of freedom which is expected to be of order $10^2$ in the early universe. At matter radiation equality, $g_{*, \rm eq}\approx3$ while $k_{\rm eq}= 0.01{\rm Mpc}^{-1}$. Hence, using \eqref{massk}, we obtain the $k$-dependence of the variance $\sigma^2(k)$.

\begin{figure}
	\includegraphics[width=0.48\textwidth]{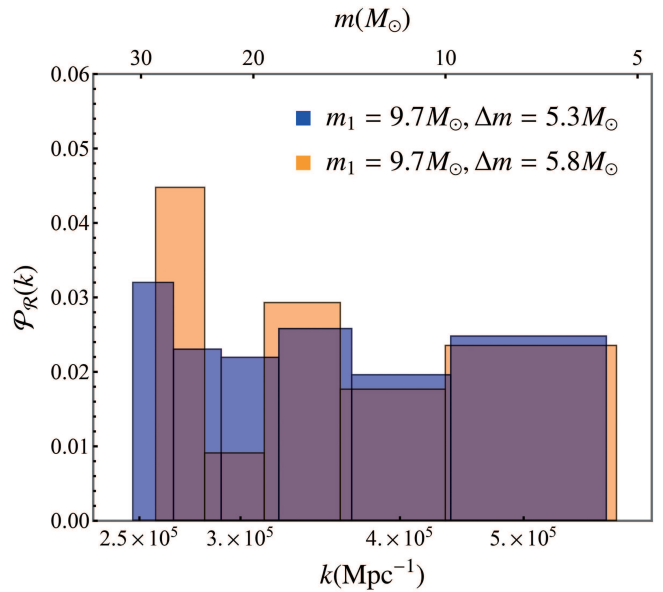}
	\caption{The power spectrum of primordial curvature perturbations derived from the reconstruction procedure. We take $\Delta_{\rm th}=0.23$ and $f_{\rm PBH}=10^{-3}$ in the calculations.} 
	\label{fig:PR} 
\end{figure}

Since $\sigma^2(k)$ is discrete, we rewrite \eqref{powersim} into the discrete form by replacing ${\rm d}\sigma^2/{\rm d}k$ with $\Delta\sigma^2/\Delta k$, then insert $\sigma^2(k)$ into it. Using the difference method, we finally reconstruct the power spectrum of primordial curvature perturbation, shown in Fig.~\ref{fig:PR}. There are 6 and 5 columns for $\Delta m=5.3M_\odot$ and $5.8M_\odot$ cases, respectively. The number of columns are reduced by one compared to the case of mass function because of the difference method we adopted. In both cases, we find that on scales $\mathcal{O}(1)$ pc, the amplitude of power spectrum is $\mathcal{O}\left(10^{-2}\right)$. This result is consistent with the PBH formation scenario that on small scales, the amplitude of power spectrum of primordial curvature perturbations is enhanced to $\mathcal{P}_{\cal R}=\mathcal{O}\left(10^{-2}\sim10^{-1}\right)$ during inflationary epoch. Especially, in case of $\Delta m=5.3M_\odot$, the power spectrum is relatively flat, while the average value of its amplitude satisfies the constraint by the effect of accretion onto the PBHs~\cite{Carr:2020gox}.

We note that in case of critical collapse where $m=K_{\rm cri}M_{\rm R}\left(\Delta-\Delta_{\rm th}\right)^\gamma$ with $K_{\rm cri}\approx3.3$ and $\gamma\approx0.36$~\cite{Choptuik:1992jv}, the relation \eqref{sigmafun1} and \eqref{massk} will be modified. Compared to the non-critical case, the same mass of PBH requires larger horizon mass. Provided with the same values of $\Delta_{\rm th}$ and $f_{\rm PBH}$, this may result in the power spectrum with relatively higher amplitude on larger scale.

We can make a consistency check of our result. In the Press-Schechter formalism, the PBH abundance $\beta$ is defined as the integration of the probability where the density contrast is larger than the threshold for PBH formation:
\begin{align}
\label{beta}
    \beta\equiv2\int_{\nu_{\rm th}}^\infty P(\nu){\rm d}\nu={\rm erfc}\left(\frac{\nu_{\rm th}}{\sqrt{2}}\right),
\end{align}
where $\nu\equiv\Delta/\sigma$. Then the PBH fraction can be expressed in terms of $\beta$ as~\cite{Carr:2020gox}
\begin{align}
\label{beta}
    f_{\rm PBH}(M)\approx&1.42\times10^{17}K^{\frac{1}{2}}\left(\frac{0.67}{h}\right)^{2}\left(\frac{106.75}{g_{*\rm{i}}}\right)^{\frac{1}{4}}\nonumber\\
    &\times\left(\frac{10^{15}{\rm g}}{M}\right)^{\frac{1}{2}}\Omega_{\rm CDM}^{-1}\beta(M)\,.
\end{align}
Using $\sigma^2(M)$ constructed above, together with $\Delta_{\rm th}=0.23$, we find $f_{\rm PBH}(M)$ takes the value in the range  $\mathcal{O}\left(10^{-3}\sim10^{-4}\right)$, which is consistent with $f_{\rm PBH}=10^{-3}$ assumed in our calculations. 

On the other hand, large linear scalar perturbations would source non-negligible tensor modes at second order. Using our result, the corresponding induced GW spectrum is roughly estimated as~\cite{Domenech:2021ztg,Pi:2021dft}
\begin{align}
\label{IGW}
    \Omega_{\rm GW}\sim10^{-6}\mathcal{P}_{\cal R}^2\sim10^{-10}\,,
\end{align}
at frequency $3\times10^{-9}$ Hz for PBHs with mass $10M_{\odot}$, which is consistent with the prediction that PBHs provide a candidate for the NANOGrav 12.5-yr signal~\cite{NANOGrav:2020bcs,Vaskonen:2020lbd,DeLuca:2020agl,Kohri:2020qqd,Yi:2021lxc}.\\


\textit{Conclusion and Discussion}~
Based on the scenario that PBHs provide a candidate for the BBH merger events detected by the LIGO/VIRGO, using the GWTC-1 to GWTC-3 catalogs, we for the first time reconstruct the power spectrum of primordial curvature perturbations on small scales. It is found that the amplitude of the power spectrum is enhanced to $\mathcal{O}\left(10^{-2}\right)$ on scales $\mathcal{O}\left(1\right)$ pc. The resulting power spectrum is consistent with the theoretical expectation of a huge enhancement of primordial curvature perturbations from $\mathcal{O}\left(10^{-5}\right)$ on CMB scale to $\mathcal{O}\left(10^{-1}\right)$ on small scales for PBH formation.

Although we have assumed the top-hat form of window function, as pointed out in \cite{Ando:2018qdb,Young:2019osy}, the effect of the choice of window function will cause the uncertainty in the amplitude of the power spectrum only up to $\mathcal{O}(10\%)$. Uncertainties mainly come from three facts: (i). The limited number of confirmed BBH merger events. Currently, there are only 91 reported BBH merger events, so we have to balance the value of mass gap $\Delta m$ and the number of events contained in the grid when we use the method developed in \eqref{f12}--\eqref{fn} in practice. As shown in Fig.~\ref{fig:grids}, the number of effective grids and events contained inside constrains the precision of the resulting power spectrum. We expect this will be improved in the future detection. (ii). Contribution of PBH mergers. Although a multi-channel with both astrophysical and PBH mergers is likely to explain the LIGO/Virgo detection~\cite{Hall:2020daa,Zevin:2020gbd,Kritos:2020wcl,Franciolini:2021tla,Bavera:2021wmw,Chen:2021nxo,Hutsi:2020sol,Wong:2020yig}, currently there is no explicit method to distinguish the  astrophysical BHs from PBHs. We simply translate the statement ``large mass" and ``low spin" into the requirements $m\gtrsim3M_{\odot}$ and $|\chi_{\rm eff}|\lesssim0.3$, respectively. In this sense, we may have over-counted the PBH merger events.  (iii). Limitation of detector. Throughout the paper, we identify the number of BBHs merger events as the ``detected" one. In reality, a detector has its selection bias; its effect is described by the detection probability $p_{\rm det}(M_1, M_2, z)$, which is the probability that a given detector detects the merger event with individual masses $M_1$ and $M_2$ occurring at redshift $z$~\cite{Chen:2017wpg,Okano:2022kio}. Our method is not sensitive to the mass dependence of $p_{\rm det}$ because with the detection of more BBH events in future, the width of grids on the  $M_1-M_2$ plane could be shortened so that the detection probabilities between neighbouring events do not vary much. However, the $z$-dependence may cause large deviation. In this sense, we have to take into account the time dependence of merger rate carefully to obtain a precise amplitude of the power spectrum.
Moreover, the primordial curvature perturbation is non-Gaussian in general~	\cite{Luo:1992er,Verde:1999ij,Verde:2000vr,Komatsu:2001rj,Bartolo:2004if,Boubekeur:2005fj,Byrnes:2007tm}. In such case, the PBH mass function and PBH abundance depend also on higher-order statistics~\cite{Bullock:1996at,Ivanov:1997ia,PinaAvelino:2005rm,Lyth:2012yp,Byrnes:2012yx,Shandera:2012ke,Young:2013oia,Franciolini:2018vbk,Cai:2018dig,Yoo:2019pma,Atal:2019erb,Ezquiaga:2019ftu,Young:2014oea,Young:2015cyn,DeLuca:2019qsy,Young:2019yug}, and a further development of the reconstruction procedure is necessary. We leave the investigation of this case for future work.

\vspace{1em}
\begin{acknowledgments}
\textit{Acknowledgments}~
We thank Kazunori Kohri, Shi Pi, Misao Sasaki and Teruaki Suyama for useful discussions. M. Y. acknowledges financial support from JSPS Grant-in-Aid for Scientific Research No. JP18K18764, JP21H01080, JP21H00069. Y. Z. is supported by the Fundamental Research Funds for the Central Universities. 
\end{acknowledgments}


\begin{thebibliography}{99}
\bibitem{Abbott:2016blz} 
  B.~P.~Abbott {\it et al.} [LIGO Scientific and Virgo Collaborations],
  Phys.\ Rev.\ Lett.\  {\bf 116}, no. 6, 061102 (2016)
  [arXiv:1602.03837 [gr-qc]].
 
\bibitem{Abbott:2016nmj} 
  B.~P.~Abbott {\it et al.} [LIGO Scientific and Virgo Collaborations],
  Phys.\ Rev.\ Lett.\  {\bf 116}, no. 24, 241103 (2016)
  [arXiv:1606.04855 [gr-qc]].
  
\bibitem{TheLIGOScientific:2016pea} 
  B.~P.~Abbott {\it et al.} [LIGO Scientific and Virgo Collaborations],
  Phys.\ Rev.\ X {\bf 6}, no. 4, 041015 (2016)
  Erratum: [Phys.\ Rev.\ X {\bf 8}, no. 3, 039903 (2018)]
  [arXiv:1606.04856 [gr-qc]].
  
\bibitem{Abbott:2017vtc} 
  B.~P.~Abbott {\it et al.} [LIGO Scientific and VIRGO Collaborations],
  Phys.\ Rev.\ Lett.\  {\bf 118}, no. 22, 221101 (2017)
  Erratum: [Phys.\ Rev.\ Lett.\  {\bf 121}, no. 12, 129901 (2018)]
  [arXiv:1706.01812 [gr-qc]].
  
\bibitem{Abbott:2017oio} 
  B.~P.~Abbott {\it et al.} [LIGO Scientific and Virgo Collaborations],
  Phys.\ Rev.\ Lett.\  {\bf 119}, no. 14, 141101 (2017)
  [arXiv:1709.09660 [gr-qc]].
  
\bibitem{Abbott:2017gyy} 
  B.~ P.~Abbott {\it et al.} [LIGO Scientific and Virgo Collaborations],
  Astrophys.\ J.\  {\bf 851}, no. 2, L35 (2017)
  [arXiv:1711.05578 [astro-ph.HE]].

\bibitem{LIGOScientific:2018mvr} 
  B.~P.~Abbott {\it et al.} [LIGO Scientific and Virgo Collaborations],
  Phys.\ Rev.\ X {\bf 9}, no. 3, 031040 (2019)
  [arXiv:1811.12907 [astro-ph.HE]].
  
\bibitem{Venumadhav:2019lyq}
T.~Venumadhav, B.~Zackay, J.~Roulet, L.~Dai and M.~Zaldarriaga,
Phys. Rev. D \textbf{101}, no.8, 083030 (2020)
[arXiv:1904.07214 [astro-ph.HE]].

\bibitem{Abbott:2020niy}
R.~Abbott \textit{et al.} [LIGO Scientific and Virgo],
[arXiv:2010.14527 [gr-qc]].

\bibitem{LIGOScientific:2021usb}
R.~Abbott \textit{et al.} [LIGO Scientific and VIRGO],
[arXiv:2108.01045 [gr-qc]].

\bibitem{LIGOScientific:2021djp}
R.~Abbott \textit{et al.} [LIGO Scientific, VIRGO and KAGRA],
[arXiv:2111.03606 [gr-qc]].
  
\bibitem{Belczynski:2010tb} 
  K.~Belczynski, M.~Dominik, T.~Bulik, R.~O'Shaughnessy, C.~Fryer and D.~E.~Holz,
  Astrophys.\ J.\  {\bf 715}, L138 (2010)
  [arXiv:1004.0386 [astro-ph.HE]].
  
\bibitem{Dominik:2012kk} 
  M.~Dominik, K.~Belczynski, C.~Fryer, D.~Holz, E.~Berti, T.~Bulik, I.~Mandel and R.~O'Shaughnessy,
  Astrophys.\ J.\  {\bf 759}, 52 (2012)
  [arXiv:1202.4901 [astro-ph.HE]].
  
\bibitem{Dominik:2013tma} 
  M.~Dominik, K.~Belczynski, C.~Fryer, D.~E.~Holz, E.~Berti, T.~Bulik, I.~Mandel and R.~O'Shaughnessy,
  Astrophys.\ J.\  {\bf 779}, 72 (2013)
  [arXiv:1308.1546 [astro-ph.HE]].
  
\bibitem{Dominik:2014yma} 
  M.~Dominik {\it et al.},
  Astrophys.\ J.\  {\bf 806}, no. 2, 263 (2015)
  [arXiv:1405.7016 [astro-ph.HE]].
  
\bibitem{Belczynski:2016obo} 
  K.~Belczynski, D.~E.~Holz, T.~Bulik and R.~O'Shaughnessy,
  Nature {\bf 534}, 512 (2016)
  [arXiv:1602.04531 [astro-ph.HE]].

\bibitem{TheLIGOScientific:2016htt} 
  B.~P.~Abbott {\it et al.} [LIGO Scientific and Virgo Collaborations],
  Astrophys.\ J.\  {\bf 818}, no. 2, L22 (2016)
  [arXiv:1602.03846 [astro-ph.HE]].
  
\bibitem{Miller:2016krr} 
  M.~Coleman Miller,
  Gen.\ Rel.\ Grav.\  {\bf 48}, no. 7, 95 (2016)
  [arXiv:1606.06526 [astro-ph.HE]].
  
\bibitem{Bird:2016dcv} 
  S.~Bird, I.~Cholis, J.~B.~Munoz, Y.~Ali-Haimoud, M.~Kamionkowski, E.~D.~Kovetz, A.~Raccanelli and A.~G.~Riess,
  Phys.\ Rev.\ Lett.\  {\bf 116}, no. 20, 201301 (2016)
  [arXiv:1603.00464 [astro-ph.CO]].

\bibitem{Clesse:2016vqa} 
  S.~Clesse and J.~Garcia-Bellido,
  Phys.\ Dark Univ.\  {\bf 15}, 142 (2017)
  [arXiv:1603.05234 [astro-ph.CO]].
  
\bibitem{Sasaki:2016jop} 
  M.~Sasaki, T.~Suyama, T.~Tanaka and S.~Yokoyama,
  Phys.\ Rev.\ Lett.\  {\bf 117}, no. 6, 061101 (2016)
  Erratum: [Phys.\ Rev.\ Lett.\  {\bf 121}, no. 5, 059901 (2018)]
  [arXiv:1603.08338 [astro-ph.CO]].
  
  \bibitem{Sasaki:2018dmp} 
  M.~Sasaki, T.~Suyama, T.~Tanaka and S.~Yokoyama,
  Class.\ Quant.\ Grav.\  {\bf 35}, no. 6, 063001 (2018)
  [arXiv:1801.05235 [astro-ph.CO]].

\bibitem{Garcia-Bellido:2020pwq}
J.~Garcia-Bellido, J.~F.~N.~Siles and E.~Ruiz Morales,
arXiv:2010.13811 [astro-ph.CO].



\bibitem{Zeldovich:1963}
Ya. B. Zel'dovich and I.D. Novikov, Sov. Astron. {\bf{10}}, 602 (1966).
	
\bibitem{Hawking:1971ei} 
S.~Hawking,
Mon.\ Not.\ Roy.\ Astron.\ Soc.\  {\bf 152}, 75 (1971).


\bibitem{Carr:1974nx} 
B.~J.~Carr and S.~W.~Hawking,
Mon.\ Not.\ Roy.\ Astron.\ Soc.\  {\bf 168}, 399 (1974).


\bibitem{Meszaros:1974tb} 
P.~Meszaros,
Astron.\ Astrophys.\  {\bf 37}, 225 (1974).

	
\bibitem{Carr:1975qj} 
B.~J.~Carr,
Astrophys.\ J.\  {\bf 201}, 1 (1975).


\bibitem{Bringmann:2011ut}
T.~Bringmann, P.~Scott and Y.~Akrami,
Phys.\ Rev.\ D {\bf 85} (2012) 125027
[arXiv:1110.2484 [astro-ph.CO]].

	
\bibitem{Green:2018akb} 
A.~M.~Green,
Phys.\ Rev.\ D {\bf 98}, no. 2, 023529 (2018)
[arXiv:1805.05178 [astro-ph.CO]].

\bibitem{Planck:2018jri}
Y.~Akrami \textit{et al.} [Planck],
Astron. Astrophys. \textbf{641}, A10 (2020)
[arXiv:1807.06211 [astro-ph.CO]].




\bibitem{Chluba:2019kpb}
J.~Chluba, A.~Kogut, S.~P.~Patil, M.~H.~Abitbol, N.~Aghanim, Y.~Ali-Ha\"\i{}moud, M.~A.~Amin, J.~Aumont, N.~Bartolo and K.~Basu, \textit{et al.}
Bull. Am. Astron. Soc. \textbf{51}, no.3, 184 (2019)
[arXiv:1903.04218 [astro-ph.CO]].

\bibitem{Jeong:2014gna}
D.~Jeong, J.~Pradler, J.~Chluba and M.~Kamionkowski,
Phys. Rev. Lett. \textbf{113}, 061301 (2014)
[arXiv:1403.3697 [astro-ph.CO]].

\bibitem{Nakama:2014vla}
T.~Nakama, T.~Suyama and J.~Yokoyama,
Phys. Rev. Lett. \textbf{113}, 061302 (2014)
[arXiv:1403.5407 [astro-ph.CO]].

\bibitem{Inomata:2016uip}
K.~Inomata, M.~Kawasaki and Y.~Tada,
Phys. Rev. D \textbf{94}, no.4, 043527 (2016)
[arXiv:1605.04646 [astro-ph.CO]].

\bibitem{Allahverdi:2020bys}
R.~Allahverdi, M.~A.~Amin, A.~Berlin, N.~Bernal, C.~T.~Byrnes, M.~Sten Delos, A.~L.~Erickcek, M.~Escudero, D.~G.~Figueroa and K.~Freese, \textit{et al.}
Open J. Astrophys. \textbf{4}, 2021
[arXiv:2006.16182 [astro-ph.CO]].

\bibitem{Gow:2020bzo}
A.~D.~Gow, C.~T.~Byrnes, P.~S.~Cole and S.~Young,
JCAP \textbf{02}, 002 (2021)
[arXiv:2008.03289 [astro-ph.CO]].

\bibitem{Franciolini:2022pav}
G.~Franciolini and A.~Urbano,
[arXiv:2207.10056 [astro-ph.CO]].

\bibitem{Franciolini:2022tfm}
G.~Franciolini, I.~Musco, P.~Pani and A.~Urbano,
[arXiv:2209.05959 [astro-ph.CO]].


\bibitem{Kimura:2021sqz}
R.~Kimura, T.~Suyama, M.~Yamaguchi and Y.~L.~Zhang,
JCAP \textbf{04}, 031 (2021)
[arXiv:2102.05280 [astro-ph.CO]].


\bibitem{Raidal:2018bbj}
M.~Raidal, C.~Spethmann, V.~Vaskonen and H.~Veerm\"ae,
JCAP \textbf{02}, 018 (2019)
[arXiv:1812.01930 [astro-ph.CO]].


\bibitem{DeLuca:2020bjf}
V.~De Luca, G.~Franciolini, P.~Pani and A.~Riotto,
JCAP \textbf{04}, 052 (2020)
[arXiv:2003.02778 [astro-ph.CO]].



	
	
  
 


\bibitem{Ali-Haimoud:2016mbv}
Y.~Ali-Ha\"\i{}moud and M.~Kamionkowski,
Phys. Rev. D \textbf{95}, no.4, 043534 (2017)
[arXiv:1612.05644 [astro-ph.CO]]

\bibitem{Poulin:2017bwe}
V.~Poulin, P.~D.~Serpico, F.~Calore, S.~Clesse and K.~Kohri,
Phys. Rev. D \textbf{96}, no.8, 083524 (2017)
[arXiv:1707.04206 [astro-ph.CO]].

\bibitem{Serpico:2020ehh}
P.~D.~Serpico, V.~Poulin, D.~Inman and K.~Kohri,
Phys. Rev. Res. \textbf{2}, no.2, 023204 (2020)
[arXiv:2002.10771 [astro-ph.CO]].







\bibitem{Musco:2004ak} 
I.~Musco, J.~C.~Miller and L.~Rezzolla,
Class.\ Quant.\ Grav.\  {\bf 22}, 1405 (2005)
[gr-qc/0412063].


\bibitem{Musco:2008hv} 
I.~Musco, J.~C.~Miller and A.~G.~Polnarev,
Class.\ Quant.\ Grav.\  {\bf 26}, 235001 (2009)
[arXiv:0811.1452 [gr-qc]].


\bibitem{Musco:2012au}
I.~Musco and J.~C.~Miller,
Class.\ Quant.\ Grav.\  {\bf 30} (2013) 145009
[arXiv:1201.2379 [gr-qc]].


\bibitem{Harada:2013epa} 
~T.~Harada, C.~M.~Yoo and K.~Kohri,
Phys.\ Rev.\ D {\bf 88}, no. 8, 084051 (2013)
Erratum: [Phys.\ Rev.\ D {\bf 89}, no. 2, 029903 (2014)]
[arXiv:1309.4201 [astro-ph.CO]].

\bibitem{Yoo:2018kvb}
C.~M.~Yoo, T.~Harada, J.~Garriga and K.~Kohri,
PTEP \textbf{2018}, no.12, 123E01 (2018)
[arXiv:1805.03946 [astro-ph.CO]].

\bibitem{Musco:2018rwt}
I.~Musco,
Phys. Rev. D \textbf{100}, no.12, 123524 (2019)
[arXiv:1809.02127 [gr-qc]].


\bibitem{Green:2004wb} 
A.~M.~Green, A.~R.~Liddle, K.~A.~Malik and M.~Sasaki,
Phys.\ Rev.\ D {\bf 70}, 041502 (2004)
[astro-ph/0403181].
 




\bibitem{Carr:2020gox}
B.~Carr, K.~Kohri, Y.~Sendouda and J.~Yokoyama,
Rept. Prog. Phys. \textbf{84}, no.11, 116902 (2021)
[arXiv:2002.12778 [astro-ph.CO]].

\bibitem{Choptuik:1992jv}
M.~W.~Choptuik,
Phys. Rev. Lett. \textbf{70}, 9-12 (1993)


\bibitem{Domenech:2021ztg}
G.~Dom\`enech,
Universe \textbf{7}, no.11, 398 (2021)
[arXiv:2109.01398 [gr-qc]].

\bibitem{Pi:2021dft}
S.~Pi and M.~Sasaki,
[arXiv:2112.12680 [astro-ph.CO]].

\bibitem{NANOGrav:2020bcs}
Z.~Arzoumanian \textit{et al.} [NANOGrav],
Astrophys. J. Lett. \textbf{905}, no.2, L34 (2020)
[arXiv:2009.04496 [astro-ph.HE]].

\bibitem{Vaskonen:2020lbd}
V.~Vaskonen and H.~Veerm\"ae,
Phys. Rev. Lett. \textbf{126}, no.5, 051303 (2021)
[arXiv:2009.07832 [astro-ph.CO]].

\bibitem{DeLuca:2020agl}
V.~De Luca, G.~Franciolini and A.~Riotto,
Phys. Rev. Lett. \textbf{126}, no.4, 041303 (2021)
[arXiv:2009.08268 [astro-ph.CO]].

\bibitem{Kohri:2020qqd}
K.~Kohri and T.~Terada,
Phys. Lett. B \textbf{813}, 136040 (2021)
[arXiv:2009.11853 [astro-ph.CO]].

\bibitem{Yi:2021lxc}
Z.~Yi and Z.~H.~Zhu,
JCAP \textbf{05}, no.05, 046 (2022)
doi:10.1088/1475-7516/2022/05/046
[arXiv:2105.01943 [gr-qc]].


\bibitem{Ando:2018qdb} 
K.~Ando, K.~Inomata and M.~Kawasaki,
Phys.\ Rev.\ D {\bf 97}, no. 10, 103528 (2018)
[arXiv:1802.06393 [astro-ph.CO]].

\bibitem{Young:2019osy} 
S.~Young,
Int.\ J.\ Mod.\ Phys.\ D {\bf 29}, no. 02, 2030002 (2019)
[arXiv:1905.01230 [astro-ph.CO]].

\bibitem{Hall:2020daa}
A.~Hall, A.~D.~Gow and C.~T.~Byrnes,
Phys. Rev. D \textbf{102}, 123524 (2020)
[arXiv:2008.13704 [astro-ph.CO]].

\bibitem{Wong:2020yig}
K.~W.~K.~Wong, G.~Franciolini, V.~De Luca, V.~Baibhav, E.~Berti, P.~Pani and A.~Riotto,
Phys. Rev. D \textbf{103}, no.2, 023026 (2021)
doi:10.1103/PhysRevD.103.023026
[arXiv:2011.01865 [gr-qc]].

\bibitem{Zevin:2020gbd}
M.~Zevin, S.~S.~Bavera, C.~P.~L.~Berry, V.~Kalogera, T.~Fragos, P.~Marchant, C.~L.~Rodriguez, F.~Antonini, D.~E.~Holz and C.~Pankow,
Astrophys. J. \textbf{910}, no.2, 152 (2021)
[arXiv:2011.10057 [astro-ph.HE]].

\bibitem{Hutsi:2020sol}
G.~H\"utsi, M.~Raidal, V.~Vaskonen and H.~Veerm\"ae,
JCAP \textbf{03}, 068 (2021)
[arXiv:2012.02786 [astro-ph.CO]].

\bibitem{Kritos:2020wcl}
K.~Kritos, V.~De Luca, G.~Franciolini, A.~Kehagias and A.~Riotto,
JCAP \textbf{05}, 039 (2021)
[arXiv:2012.03585 [gr-qc]].

\bibitem{Franciolini:2021tla}
G.~Franciolini, V.~Baibhav, V.~De Luca, K.~K.~Y.~Ng, K.~W.~K.~Wong, E.~Berti, P.~Pani, A.~Riotto and S.~Vitale,
Phys. Rev. D \textbf{105}, no.8, 083526 (2022)
[arXiv:2105.03349 [gr-qc]].

\bibitem{Bavera:2021wmw}
S.~S.~Bavera, G.~Franciolini, G.~Cusin, A.~Riotto, M.~Zevin and T.~Fragos,
Astron. Astrophys. \textbf{660}, A26 (2022)
[arXiv:2109.05836 [astro-ph.CO]].

\bibitem{Chen:2021nxo}
Z.~C.~Chen, C.~Yuan and Q.~G.~Huang,
Phys. Lett. B \textbf{829}, 137040 (2022)
[arXiv:2108.11740 [astro-ph.CO]].


\bibitem{Chen:2017wpg}
H.~Y.~Chen, D.~E.~Holz, J.~Miller, M.~Evans, S.~Vitale and J.~Creighton,
Class. Quant. Grav. \textbf{38}, no.5, 055010 (2021)
[arXiv:1709.08079 [astro-ph.CO]].
	
 
\bibitem{Okano:2022kio}
S.~Okano and T.~Suyama,
[arXiv:2201.10258 [astro-ph.CO]].



	

\bibitem{Luo:1992er} 
X.~c.~Luo and D.~N.~Schramm,
Astrophys.\ J.\  {\bf 408}, 33 (1993).

\bibitem{Verde:1999ij} 
L.~Verde, L.~M.~Wang, A.~Heavens and M.~Kamionkowski,
Mon.\ Not.\ Roy.\ Astron.\ Soc.\  {\bf 313}, L141 (2000)
[astro-ph/9906301].

\bibitem{Verde:2000vr} 
L.~Verde, R.~Jimenez, M.~Kamionkowski and S.~Matarrese,
Mon.\ Not.\ Roy.\ Astron.\ Soc.\  {\bf 325}, 412 (2001)
[astro-ph/0011180].
		
	\bibitem{Komatsu:2001rj} 
	E.~Komatsu and D.~N.~Spergel,
	Phys.\ Rev.\ D {\bf 63}, 063002 (2001)
	[astro-ph/0005036].
	
		
\bibitem{Bartolo:2004if} 
N.~Bartolo, E.~Komatsu, S.~Matarrese and A.~Riotto,
Phys.\ Rept.\  {\bf 402}, 103 (2004)
[astro-ph/0406398].
	
	

	\bibitem{Boubekeur:2005fj} 
	L.~Boubekeur and D.~H.~Lyth,
	Phys.\ Rev.\ D {\bf 73}, 021301 (2006)
	[astro-ph/0504046].
	
	\bibitem{Byrnes:2007tm} 
	C.~T.~Byrnes, K.~Koyama, M.~Sasaki and D.~Wands,
	JCAP {\bf 0711}, 027 (2007)
	[arXiv:0705.4096 [hep-th]].
	
	
	
	
	

	



	
	





\bibitem{Bullock:1996at} 
  J.~S.~Bullock and J.~R.~Primack,
  Phys.\ Rev.\ D {\bf 55}, 7423 (1997)
  [astro-ph/9611106].
  
  \bibitem{Ivanov:1997ia} 
  P.~Ivanov,
  Phys.\ Rev.\ D {\bf 57}, 7145 (1998)
  [astro-ph/9708224].

\bibitem{PinaAvelino:2005rm}
P.~Pina Avelino,
Phys. Rev. D \textbf{72}, 124004 (2005)
[arXiv:astro-ph/0510052 [astro-ph]].  

\bibitem{Lyth:2012yp}
D.~H.~Lyth,
JCAP \textbf{05}, 022 (2012)
[arXiv:1201.4312 [astro-ph.CO]].

\bibitem{Byrnes:2012yx}
C.~T.~Byrnes, E.~J.~Copeland and A.~M.~Green,
Phys. Rev. D \textbf{86}, 043512 (2012)
[arXiv:1206.4188 [astro-ph.CO]].

  
\bibitem{Shandera:2012ke} 
  S.~Shandera, A.~L.~Erickcek, P.~Scott and J.~Y.~Galarza,
  Phys.\ Rev.\ D {\bf 88}, no. 10, 103506 (2013)
  [arXiv:1211.7361 [astro-ph.CO]].
  
\bibitem{Young:2013oia} 
  S.~Young and C.~T.~Byrnes,
  JCAP {\bf 1308}, 052 (2013)
  [arXiv:1307.4995 [astro-ph.CO]].

\bibitem{Young:2014oea}
S.~Young and C.~T.~Byrnes,
Phys. Rev. D \textbf{91}, no.8, 083521 (2015)
[arXiv:1411.4620 [astro-ph.CO]].

\bibitem{Young:2015cyn}
S.~Young, D.~Regan and C.~T.~Byrnes,
JCAP \textbf{02}, 029 (2016)
[arXiv:1512.07224 [astro-ph.CO]].

\bibitem{Franciolini:2018vbk}
G.~Franciolini, A.~Kehagias, S.~Matarrese and A.~Riotto,
JCAP \textbf{03} (2018), 016
[arXiv:1801.09415 [astro-ph.CO]].

\bibitem{Cai:2018dig}
R.~g.~Cai, S.~Pi and M.~Sasaki,
Phys. Rev. Lett. \textbf{122}, no.20, 201101 (2019)
[arXiv:1810.11000 [astro-ph.CO]].

\bibitem{DeLuca:2019qsy}
V.~De Luca, G.~Franciolini, A.~Kehagias, M.~Peloso, A.~Riotto and C.~Unal,
JCAP \textbf{07} (2019), 048
[arXiv:1904.00970 [astro-ph.CO]].

\bibitem{Young:2019yug}
S.~Young, I.~Musco and C.~T.~Byrnes,
JCAP \textbf{11}, 012 (2019)
[arXiv:1904.00984 [astro-ph.CO]].

\bibitem{Atal:2019erb}
V.~Atal, J.~Cid, A.~Escriva and J.~Garriga,
JCAP \textbf{05}, 022 (2020)
[arXiv:1908.11357 [astro-ph.CO]].

\bibitem{Yoo:2019pma}
C.~M.~Yoo, J.~O.~Gong and S.~Yokoyama,
JCAP \textbf{09}, 033 (2019)
[arXiv:1906.06790 [astro-ph.CO]].

\bibitem{Ezquiaga:2019ftu}
J.~M.~Ezquiaga, J.~Garca-Bellido and V.~Vennin,
JCAP \textbf{03}, 029 (2020)
[arXiv:1912.05399 [astro-ph.CO]].
  
  \end{thebibliography}
\end{document}